\documentstyle[12pt]{article}
\def\a{\alpha}
\def\b{\beta}

\def\d{\delta}

\def\p{\psi}

\def\l{\lambda}

\def\o{\omega}
\def\t{\theta}
\def\r{\rho}
\def\s{\sigma}

\def\L{\Lambda}

\def\da{\dot\alpha}
\def\db{\dot\beta}

\def\br{\bar{\r}}

\def\be{\begin{equation}}
\def\ee{\end{equation}}
\def\arr{\begin{array}{rll}}
\def\ea{\end{array}}
\def\bea{\begin{eqnarray}}
\def\eea{\end{eqnarray}}

\begin{document}
\begin{titlepage}
\noindent

\vskip 3.0cm

\begin{center}

{\Large\bf CURING THE INFINITE GHOST TOWER}\\

\bigskip

{\Large\bf IN 4D SIEGEL SUPERPARTICLE}\\

\bigskip

\vskip 2cm

{\Large 
Stefano Bellucci\footnote{E-mail: Stefano.Bellucci@lnf.infn.it}
and Anton Galajinsky\footnote{E-mail: Anton.Galajinsky@lnf.infn.it}}

\vskip 1.0cm

{\it INFN--Laboratori Nazionali di Frascati, C.P. 13, 
00044 Frascati, Italia}\\

\end{center}
\vskip 1cm

\begin{abstract}
On an explicit example of the Siegel superparticle we study an 
alternative to the harmonic superspace approach. The latter seems to be 
the only method  for quantizing infinitely reducible first class constraints 
currently available. In an appropriately extended phase space, the infinite
ghost tower is effectively canceled by that coming from the sector of 
auxiliary variables. After a proper $BRST$ treatment the theory proves to 
be of rank two which correlates well with the results obtained earlier 
within the framework of the harmonic superspace approach. The advantage of 
the novel technique, however, is the existence of an explicit Lagrangian 
formulation and the standard spin--statistics relations which hold for all 
the variables involved. 
\end{abstract}

\vspace{0.5cm}

PACS: 04.60.Ds; 11.30.Pb\\
Keywords: BRST Quantization, Supersymmetric Effective Theories

\end{titlepage}

\noindent
{\bf 1. Introduction}\\

The superparticle due to Siegel~\cite{s1} (later referred to as $AB$ superparticle)
has originally been proposed as a way to avoid problematic second class constraints
intrinsic to a conventional superparticle (superstring) theory 
without sacrificing manifest Lorentz covariance. To compensate a mismatch~\cite{all} 
in the number of degrees of freedom between the $AB$ model and the conventional 
formulation~\cite{brink}, it was suggested~\cite{s2}
to introduce  further ($C$ and $D$) constraints, the total set forming a closed 
algebra. The equivalence of the $ABCD$ theory to the
conventional superparticle (superstring) has been claimed in Ref.~\cite{miksig},
thus suggesting an intriguing alternative to the standard formalism\footnote{It 
has to be mentioned, however, that only in the string case the proof seems to be completely
consistent. For the mechanics analogue the higher order fermionic $C$ constraints were treated  
differently ({\it a la} Gupta--Bleuler) from others.}. Yet, although in the modified
theories constraints do form a closed algebra and are straightforward to
be realized quantum mechanically (operator quantization)~\cite{g1}, the 
results of the path integral quantization~\cite{kal}--\cite{gr} seem to 
be intractable because of infinite reducibility of the fermionic 
constraints involved.

A common  way to attack the latter problem, which proved to be
successful for the original  superparticle and 
superstring~\cite{nps}--\cite{ban} (field theory applications were proposed 
earlier in Ref.~\cite{iv}), is to make use of Lorentz harmonics to extract linearly independent 
components from the fermionic constraints in a covariant fashion. 
For the $AB$ model this has been accomplished in Ref.~\cite{nis}
yielding a theory of rank two after a proper $BRST$ treatment. It has to
be noted, however, that the approach of Ref.~\cite{nis} is essentially
Hamiltonian. Moreover, the standard spin--statistics relations do not hold for
some of the variables involved. Another serious problem is the noncompactness
of the coset space parametrized by the harmonics used (see the 
discussion in Ref.~\cite{del}).

Recently, an alternative technique to cure the infinite ghost tower problem
intrinsic to the Siegel superparticle, superstring has been proposed in
Ref.~\cite{sup}. The idea was to appropriately extend the original phase space and then 
effectively cancel the infinite ghost tower by that coming from the sector of auxiliary 
variables.

In the present paper we investigate in full details this, looking somewhat exotic, 
possibility and show that the result of the quantization correlates well with that 
obtained previously within the framework of the harmonic superspace approach~\cite{nis}.
The advantage of the novel scheme, however, is the presence of an explicit 
Lagrangian formulation and the standard spin--statistic relations which hold for all the variables. 
For simplicity of the presentation we restrict ourselves to the four 
dimensional case. With some modifications, however, this can be generalized to other 
dimensions~\cite{sup}. 

In the next section we review the $AB$ model in $d=4$. It is shown that, in contrast to 
the common opinion, a ghost free and unitary quantum mechanics can be constructed if 
one sacrifices  conventional conjugation properties for fermionic operators
and chooses a specific modification. A light--cone Hilbert space is 
explicitly constructed which can be identified with the one particle sector of
the quantized supersymmetric massless Wess--Zumino model. In Sec. 3, following the ideology of an 
earlier work~\cite{sup}, we embed the original Siegel model into an appropriately
extended configuration space. In the Hamiltonian framework, the variables
from the auxiliary sector turn out to be subjected to {\it reducible} constraints like those 
entering Siegel's theory. These can further be used to put the complete constraint set into
the form of {\it first} stage of reducibility, the latter admitting straightforward 
path integral quantization~\cite{bat1}. Sec. 4 contains the construction of the 
$BRST$ charge in the minimal
ghost sector. The theory in the extended phase space proves to be rank two. This
is in perfect agreement with the analysis of the alternative harmonic superspace
approach. An extension to the nonminimal ghost sector and the path integral quantization
are accomplished in Sec. 5. We conclude with some remarks on possible further developments
of the formalism in Sec. 6. Appendix contains the light--cone notation and technical points
related to the Hamiltonian analysis of Sec. 3. Throughout the paper we use the spinor notation 
from Ref.~\cite{kuz}.

\vspace{0.4cm}

\noindent
{\bf 2. Review of the $4d$ $AB$ model. No negative norm states and
unitarity in the physical subspace.}\\

Retaining only first class constraints of the conventional 
superparticle\footnote{Conventions adopted in this section 
are ${(\t^\a)}^{*}={\bar\t}^{\da}$,
${(p_{\t\a})}^{*}=-p_{\bar\t \da}$.}
(see Ref.~\cite{g1} for the details of the Dirac procedure)
\be\label{sconst}
p^2=0, \qquad {(p_\t \s^n p_n)}_{\da}=0, \qquad {(\s^n p_{\bar\t} p_n)}_\a=0,
\ee
where $(p_n,p_{\t\a},p_{\bar\t {\da}})$ are momenta conjugate to the 
configuration space variables $(x^n,\t^\a,{\bar\t}^{\da})$, one discovers 
the constraint set to describe Siegel's superparticle~\cite{s1}. 
The corresponding canonical Hamiltonian reads
\be
H={\textstyle{\frac 12}}{p^2}.
\ee

Owing to the null vector $p_n$ entering the problem, only half of
the fermionic constraints is linearly independent. 
In particular, the identity
\begin{equation}\label{pr}
{(p_{\theta} \sigma^n p_n)}_{\dot\alpha} {Z_1}^{\dot\alpha \alpha}
+{Z_1}^\alpha p^2\equiv0,
\end{equation}
where ${Z_1}^{\dot\alpha \alpha}={({\tilde\sigma}^n p_n)}^{\dot\alpha 
\alpha}, {Z_1}^\beta={p_\theta}^\beta$, is satisfied. 
On the constraint surface not all of the functions 
${Z_1}^{\dot\alpha \alpha}$ prove to be independent 
\begin{equation}\label{pr}
{Z_1}^{\dot\alpha \alpha}{Z_2}_{\alpha\dot\beta}\approx0,\quad
{Z_2}_{\alpha\dot\beta}={(\sigma^n p_n)}_{\alpha\dot\beta}.
\end{equation}
Apparently, this process can be continued, the system at hand being  
infinite stage of reducibility following the terminology of 
Ref.~\cite{bat1}.

Proceeding to the light--cone analysis of the model, one imposes the 
conventional gauge in the fermionic sector ($A^{\pm}= 
\pm{\textstyle{\frac {1}{\sqrt{2}}}}(A^{0}\pm A^{3})$)
\be
\theta\sigma^+=0, \qquad  \sigma^+\bar\theta=0,
\ee
or, reducing this to components
\be
\theta^2=0, \qquad {\bar\theta}^{\dot2}=0.
\ee
The partially reduced phase space includes then three pairs\footnote{In 
what follows we omit the indices carried by the Fermi variables.}
$(x^n,p_n)$,$(\t,p_\t)$,$(\bar\t,p_{\bar\t})$, these obeying usual 
canonical commutation relations and the conjugation 
properties ($x^n$, $p_n$ are real)
\be\label{conj}
{(\t)}^{*}=\bar\t, \qquad {(p_\t)}^{*}=
-p_{\bar\t}.
\ee
The gauge fixed Hamiltonian action acquires the form
\be\label{gfa}
S=\int d\tau \{ p_m\dot x^m+p_\theta \dot\theta+p_{\bar\theta}
\dot{\bar\theta}-\frac 12 p^2 \}.
\ee

Going over to the quantum description $(\{ \hat\t,\hat{p_\t} \}=i,
\{ \hat{\bar\t},\hat{p_{\bar\t}} \}=i)$, it is customary to require a scalar 
product in a Hilbert space to respect the conjugation property~(\ref{conj}),
i.e.
\be\label{qconj}
{\hat\t}^{+}=\hat{\bar\t}, \qquad {\hat{p_\t}}^{+}=-\hat{p_{\bar\t}}.
\ee
This, however, immediately leads us to the conclusion that there are negative 
norm states in such a quantum space. Actually, introducing the 
operators
\bea\label{aabb}
&& \hat a={\textstyle{\frac {1}{\sqrt {2}}}}(\hat \t-i{\hat p}_{\bar\t}), 
\qquad {\hat a}^{+}={\textstyle{\frac {1}{\sqrt {2}}}}
(\hat {\bar\t}-i{\hat p}_\t),
\nonumber \\
&& \qquad \qquad \qquad \{\hat a, {\hat a}^{+}\}=1, \nonumber \\
&& \hat b={\textstyle{\frac {1}{\sqrt {2}}}}(\hat \t+i{\hat p}_{\bar\t}), 
\qquad {\hat b}^{+}={\textstyle{\frac {1}{\sqrt {2}}}}
(\hat {\bar\t}+i{\hat p}_\t),
\nonumber \\
&& \qquad \qquad \qquad \{\hat b, {\hat b}^{+}\}=-1,
\eea
with a representation space being a tensor product of the corresponding 
Fock spaces, one discovers a ghost state due to the last line in 
Eq.~(\ref{aabb}).

It does not seem to have been emphasized previously, that a ghost free
quantum mechanics still can be constructed if we sacrifice 
Eq.~(\ref{qconj}) and choose the alternative
\be\label{conjmod}
{{\hat p}_\t}^{+}=-i\hat\t, \qquad {{\hat p}_{\bar\t}}^{+}=-i\hat{\bar\t}.
\ee
With such a choice, the operators
\bea\label{aabb}
&& \hat a={\textstyle{\frac {1}{\sqrt {2}}}}(\hat \t-i{\hat p}_{\bar\t}), 
\qquad {\hat a}^{+}={\textstyle{\frac {1}{\sqrt {2}}}}
(\hat {\bar\t}-i{\hat p}_\t),
\nonumber \\
&& \hat b={\textstyle{\frac {1}{\sqrt {2}}}}(\hat \t+i{\hat p}_{\bar\t}), 
\qquad {\hat b}^{+}=-{\textstyle{\frac {1}{\sqrt {2}}}}
(\hat {\bar\t}+i{\hat p}_\t),
\eea
obey
\be
\{\hat a, {\hat a}^{+}\}=1, \qquad \{\hat b,{\hat b}^{+}\}=1,
\ee
and the corresponding Fock space, obviously, does not involve ghosts.
It is worth mentioning that, there is no any physical obstruction to 
define a conjugation as in Eq. (\ref{conjmod}) because eigenvalues of the 
Fermi operators are odd supernumbers and do not correspond to any 
physical observables. Notice also that the gauge fixed action~(\ref{gfa})
remains to be real under the modified conjugation~(\ref{conjmod}),
provided the integration by parts has been performed (one can easily check 
that the variation problem is not influenced by the conjugation
since the fermions satisfy the first order free equations).

An explicit representation of the operators $(\hat\t, \hat{\bar\t},
{\hat p}_\t,{\hat p}_{\bar\t})$
in a quantum space with a scalar product respecting Eq.~(\ref{conjmod})
has been given in Ref.~\cite{g1} (similar issues have been discussed in
 Ref.~\cite{wh1}). This is realized on a linear span of the four vectors 
$(\left|0\right\rangle, 
\left|\uparrow \right\rangle, \left|\downarrow\right\rangle,
\left|\uparrow \downarrow\right\rangle)$, which we collectively call
$\left|\s \right\rangle$, with $(\hat\t,\hat{\bar\t},{\hat p}_\t,{\hat p}_{\bar\t})$ operating like
\bea
&&
\hat\theta\left|0\right\rangle=0, \quad \hat\theta\left|\uparrow\right\rangle
=i\left|0\right\rangle, \quad \hat\theta\left|\downarrow\right\rangle=0,
\quad \hat\theta\left|\uparrow \downarrow\right\rangle=
i\left|\downarrow\right\rangle,
\nonumber \\
&&
\hat{\bar\theta}\left|0\right\rangle=0, \quad 
\hat{\bar\theta}\left|\uparrow \right\rangle=0,
\quad \hat{\bar\theta}\left|\downarrow\right\rangle=i\left|0\right\rangle,
\quad \hat{\bar\theta}\left|\uparrow \downarrow \right\rangle=
-i\left|\uparrow\right\rangle,
\nonumber \\
&&
\hat {p_\theta}\left|0\right\rangle=\left|\uparrow\right\rangle, \quad 
\hat {p_\theta}\left|\uparrow\right\rangle=0, \quad
\hat {p_\theta}\left|\downarrow\right\rangle=\left|\uparrow\downarrow
\right\rangle,\quad
\hat {p_\theta}\left|\uparrow\downarrow\right\rangle=0, \nonumber \\
&&
\hat {p_{\bar\theta}}\left|0\right\rangle=\left|\downarrow\right\rangle, \quad 
\hat {p_{\bar\theta}}\left|\uparrow\right\rangle=-\left|\uparrow\downarrow
\right\rangle, \quad
\hat {p_{\bar\theta}}\left|\downarrow\right\rangle=0, \quad    
\hat {p_{\bar\theta}}\left|\uparrow\downarrow \right\rangle=0,
\eea
and
\be
\langle \s|\s'\rangle=\delta_{\s\s'}.
\ee

The total Hilbert space is defined to be a tensor product of the linear span
and the space of square integrable functions on which ${\hat x}^n$ and
${\hat p}_n$ act in the usual coordinate representation. 

A physical Hilbert space in the complete quantum space is specified
by the only constraint remaining
\begin{equation}
\hat p^2|{\rm phys}\rangle=0.
\end{equation}
Restricting ourselves to momentum eigenfunction, one finds
\bea\label{mf}
\Phi_{p,\s}={\textstyle{\frac {1}{\sqrt{2 {(2\pi)}^3 }}}}
\left|\s \right\rangle \otimes e^{-ip^{0} t + i\vec{p}\vec{x}},
\eea
where for physical reasons we have chosen an upper shell of the 
light cone $p^{0}=\sqrt{ {\vec{p}}^{2}}$.

A scalar product in the physical subspace is given by
\be
\langle \Phi|\Psi\rangle=i\int {d^3} \vec{x}
(\bar\Phi \partial_{0} \Psi-\partial_{0} \bar\Phi \Psi),
\ee
or
\be
\langle \Phi_{p,\s}| \Phi_{p',\s'} \rangle=p^0 
\delta^{(3)}(\vec{p}-\vec{p}^{'}) \delta_{\s \s'},
\ee
for the momentum eigenfunctions.

It is instructive then to clarify the structure of the Pauli-Lubanski 
vector for the case at hand. Putting the classical expression
\begin{equation}
W_a=\frac12\epsilon_{abcd}p^bS^{cd}, \\
\end{equation}
with
$S^{cd}={\t^\delta} {{\left(\sigma^{cd}\right)}_\delta}^\gamma
 p_{\t\gamma}+p_{\bar\t \dot\gamma}
{{\left({\tilde\sigma}^{cd}\right)}^{\dot\gamma}}_{\dot\delta} 
{\bar\theta}^{\dot\delta}$ being the spin part of the Lorentz generators, onto the
surface of the constraints and gauges, one obtains
\begin{equation}\label{pl}
{W_a}=\frac i2 p_a\left(p_{\theta}\theta-p_{\bar\theta}\bar\theta\right). 
\end{equation}
Here we made use of the identities ($\epsilon_{0123}=1$)
\begin{equation}
\sigma_{ab}=-\frac i2 \epsilon_{abcd}\sigma^{cd}, \qquad
{\tilde\sigma}_{ab}=\frac i2 \epsilon_{abcd}{\tilde\sigma}^{cd}.
\end{equation}
Owing to the minus sign between the two terms entering Eq.~(\ref{pl}),
one does not face any operator ordering ambiguities in passing to quantum 
description. In particular,
\be
{\hat W}_a \Phi_{p,\s}=\s p_a  \Phi_{p,\s},
\ee
where the number coefficient $\s$ takes values 
\be\label{hel}
\s=\textstyle{(0,-\frac 12,\frac 12,0)},
\ee
for the states $\left|\s \right\rangle=(\left|0\right\rangle, 
\left|\uparrow \right\rangle, \left|\downarrow\right\rangle,
\left|\uparrow \downarrow\right\rangle)$, respectively. Observe also
that ${\hat W}_a$ is hermitian with respect to both
conjugation prescriptions~(\ref{qconj}),(\ref{conjmod}).

Since the construction of unitary irreducible representations (irreps) 
of the Poincar\'e group reduces to that of the
little group generated by ${\hat W}_a$ (see e.g. \cite{wein}), it is 
straightforward to verify that given the vector $\left|\s \right\rangle
=(\left|0\right\rangle, \left|\uparrow \right\rangle, 
\left|\downarrow\right\rangle,\left|\uparrow \downarrow\right\rangle)$ 
in Eq.~(\ref{mf}) the corresponding linear space ($p_a$ takes values on the
upper shell of the light cone) realizes a unitary irrep of helicity $\s$, 
with $\s$ being specified in Eq.~(\ref{hel}).

Finally, it is worth mentioning, that the set of helicities obtained
allows us to identify the quantum space constructed with the one particle
sector of a quantized supersymmetric massless Wess--Zumino model. This correlates 
well with the results of the Dirac quantization\footnote{In Ref.~\cite{g1}
quantum wave functions were realized on real scalar superfields.
This condition can be weakened to include complex scalar superfields
if one makes proper use of both of the equation entering Eq.~(34b)
of Ref.~\cite{g1}.} accomplished in Ref.~\cite{g1}.

A path integral representation for the superpropagator of the massless 
Wess--Zumino model that explicitly involves a gauge fixed action of the $4d$
Siegel superparticle has been given in Ref.~\cite{g1}.

\vspace{0.4cm}

\noindent
{\bf 3. Siegel superparticle in an extended phase space}\\

\vspace{0.4cm}

As was demonstrated in the previous section, the Siegel superparticle
in the original formulation is infinite stage of reducibility.
In this section we reformulate the  model by
introducing a set of auxiliary variables. The extension makes it
possible to put fermionic constraints into an irreducible form 
valid for subsequent path integral quantization. 

\vspace{0.4cm} 

\noindent
{\bf 3.1. Action and symmetries}

\vspace{0.4cm} 

The superparticle action to be analyzed is
\bea
S=\int d\tau \{ \frac 1{2e} 
{(\dot x^m+i\t\s^m\dot{\bar\t}-
i\dot\t\s^m\bar\t+i\p\s^m\br-i\r\s^m\bar\p+\o\L^m)}^2
\nonumber\\[2pt]  
-\ \r^\a{\dot\t}_\a-{\br}_{\dot\a}{\dot{\bar\t}}^{\dot\a}-\o
-\phi\L^2-\L_m i\varphi\s^m\bar\chi+\L_m i\chi\s^m\bar\varphi \}.
\eea

The theory is invariant under the standard rigid supersymmetry 
transformations. The local reparametrizations and $\kappa$-symmetry
of Siegel's model
$$
\begin{array}{lll} \delta_\alpha\theta=\alpha\dot\theta, &
\delta_\alpha\bar\theta=\alpha\dot{\bar\theta}, &
\delta_\alpha x^n=\alpha\dot x^n,\\
\delta_\alpha\rho=\alpha\dot\rho, &
\delta_\alpha\bar\rho=\alpha\dot{\bar\rho}, &
\delta_\alpha e=(\alpha e)^\cdot,\\
\delta_\alpha\psi=(\alpha\psi)^\cdot, &
\delta_\alpha\bar\psi=(\alpha\bar\psi)^\cdot, &
\delta_\alpha \o=(\alpha \o)^\cdot,\\
\delta_\alpha \L^n=\alpha\dot \L^n, &
\delta_\alpha\chi=\alpha\dot\chi, &
\delta_\alpha\bar\chi=\alpha\dot{\bar\chi},\\
\delta_\alpha\varphi=(\alpha\varphi)^\cdot, &
\delta_\alpha\bar\varphi=(\alpha\bar\varphi)^\cdot, &
\delta_\alpha\phi=(\alpha\phi)^\cdot, 
\end{array}
$$
$$
\begin{array}{l}
\delta_\kappa\theta=\displaystyle -ie^{-1} \Pi_n\sigma^n\bar\kappa,
\qquad \delta_\kappa\bar\theta= ie^{-1} \Pi_n\kappa\sigma^n,\\
\delta_\kappa x^n=i\delta_\kappa \theta\sigma^n\bar\theta
-i\theta\sigma^n \delta_\kappa \bar\theta-i\kappa\sigma^n\bar\rho+
i\rho\sigma^n\bar\kappa,\\
\delta_\kappa e=4\dot\theta\kappa+4\bar\kappa\dot{\bar\theta}, \qquad
\delta_\kappa\psi=\dot\kappa,\\
\delta_\kappa\bar\psi=\dot{\bar\kappa},\end{array}
\eqno{(26)}
$$
\addtocounter{equation}{1}
where $\Pi^m=\dot x^m+i\t\s^m\dot{\bar\t}-
i\dot\t\s^m\bar\t+i\p\s^m\br-i\r\s^m\bar\p+\o\L^m$, are extended 
by two new symmetries depending on fermionic parameters $\b,\gamma$, the latter 
acting in the sector of the new variables
\be
\delta_\b \chi=\bar\b {\tilde\s}^n \L_n, \quad
\delta_\b \bar\chi=\L_n {\tilde\s}^n \b, \quad
\delta_\b \phi=i(\varphi\b-\bar\varphi \bar\b),
\ee
\be
\delta_\gamma \varphi=\bar\gamma {\tilde\s}^n \L_n, \quad
\delta_\gamma \bar\varphi=\L_n {\tilde\s}^n \gamma, \quad
\delta_\gamma\phi=-i(\chi\gamma-\bar\chi \bar\gamma).
\ee

From the transformation rules above, one concludes that the variables 
$(x^m,\t^\a,{\bar\t}_{\da})$
parametrize a conventional ${R}^{4|4}$ superspace, 
$(e,\p^\a,{\bar\p}_{\da})$ prove to be gauge fields for local
reparametrizations and $\kappa$--symmetry, whereas the pair
$(\r^\a, {\bar\r}_{\da})$ provides the terms corresponding to 
a (mixed) covariant propagator for fermions. This holds as
in the Siegel model. As shown below, there
is no dynamics in the sector of the new variables
$(\o,\L^m,\phi,\varphi^\a,{\bar\varphi}_{\da},\chi^\a,
{\bar\chi}_{\da})$, these  prove to be  purely auxiliary.

\vspace{0.4cm}

\noindent
{\bf 3.2. Fermionic constraints made irreducible}

\vspace{0.4cm}

Proceeding to the Hamiltonian analysis one finds fourteen primary 
constraints\footnote{We define momenta conjugate to Fermi 
variables to be right derivatives of a Lagrangian with respect to  
velocities. This corresponds to the following choice of the Poisson
brackets $\{\theta^\a,{p_\t}_\b\}=
{\delta^\a}_\b, \{{\bar\theta}_{\dot\alpha},
{p_{\bar\theta}}^{\dot\beta}\}=
{\delta_{\dot\alpha}}^{\dot\beta}$ and the position of momenta and
velocities in the Hamiltonian as specified below in Eq. (\ref{H}).
Our conventions for the conjugation slightly differ from those used in the 
review section  ${(\t^\a)}^{*}={\bar\t}^{\da}$,
${(p_{\t\a})}^{*}=p_{\bar\t \da}$.}
\begin{eqnarray}\label{primary} 
&&p_e=0, \quad p_\p=0, \quad p_{\bar\p}=0, \quad p_\r=0, 
\quad p_{\br}=0, \quad p_\o=0, \nonumber \\
&&p_\L=0, \quad p_\phi=0, \quad p_{\varphi}=0, \quad 
p_{\bar\varphi}=0,\quad p_{\chi}=0, \quad p_{\bar\chi}=0,
\nonumber\\
&&p_{\t\a}-p_n i{(\s^n\bar\t)}_\a-
\r_\a=0, \quad
{p_{\bar\t}}^{\da}+p_n i{(\t{\s}^n)}^{\da}-{\br}^
{\da}=0, 
\end{eqnarray}
where $p_q$ stands for a momentum canonically 
conjugate to a variable $q$.
The total Hamiltonian has the form
\begin{eqnarray}\label{H}
&& H=p_e\lambda_e+p_{\p\a}{\lambda_\p}^\alpha
+{p_{\bar\psi}}^{\dot\alpha}\lambda_{{\bar\psi}\dot\alpha}
+p_{\rho\alpha}{\lambda_\rho}^\a+{p_{\bar\rho}}^{\da}
\lambda_{{\bar\rho}\dot\alpha}
+p_\o \l_\o+p_{\L n} {\l_\L}^n+p_\phi \l_\phi \nonumber \\
&&+p_{\varphi\a} {\l_\varphi}^\a+{p_{\bar\varphi}}^{\da} \l_{{\bar\varphi}\da}+
p_{\chi\a}{\l_\chi}^\a+{p_{\bar\chi}}^{\da} \l_{{\bar\chi}\da}
+{(p_{\bar\theta}+p_n i\theta\sigma^n-\bar\rho)}^{\dot\alpha}
\lambda_{\bar\theta\dot\alpha} \nonumber \\
&&+{(p_\theta-p_n i\sigma^n\bar\theta-\rho)}
_\alpha{\lambda_\theta}^\alpha
+{\textstyle{\frac 12}} e p^2-i\psi \sigma^n \bar\rho p_n
+i\rho \sigma^n \bar\psi p_n+\phi {\L}^2+\o(1-\L p) \nonumber \\
&&+
i\varphi \sigma^n \bar\chi \L_n-i\chi \sigma^n\bar\varphi \L_n,
\end{eqnarray}
where $\lambda_{\dots}$ are the Lagrange multipliers 
associated to the primary constraints. The conservation in time
of the primary constraints yields the secondary ones
\bea\label{secondary}
&& p^2=0, \qquad p_n{(\sigma^n\bar\rho)}_\a=0, 
\qquad p_n{(\rho\sigma^n)}_{\da}=0,\nonumber \\
&& \L_n{(\sigma^n\bar\chi)}_\a=0, 
\qquad \L_n{(\chi\sigma^n)}_{\da}=0,\nonumber \\
&& \L_n{(\sigma^n\bar\varphi)}_\a=0, 
\qquad \L_n{(\varphi\sigma^n)}_{\da}=0,\nonumber \\
&& \L^2=0, \qquad 1-\L p=0,\nonumber \\
&&-2\phi\L^n+\o p^n-i\varphi \sigma^n \bar\chi
+i\chi \sigma^n\bar\varphi=0,
\eea
and fixes some of the Lagrange multipliers,
\begin{equation}\label{lm}
\begin{array}{ll} \lambda_\theta=-p_n i\sigma^n\bar\psi,& \qquad
\lambda_{\bar\theta}=i\psi\sigma^n p_n,\\
\lambda_\rho=-2p_n i\sigma^n\lambda_{\bar\theta}\approx0,& \qquad
\lambda_{\bar\rho}=2i\lambda_\theta\sigma^n p_n\approx0.
\end{array}
\end{equation}

Beautifully enough, the last equation in Eq.~(\ref{secondary})
can be simplified to (a proof is given in Appendix)
\be\label{o}
\o=0, \qquad -2\phi-i\varphi \sigma^n \bar\chi p_n
+i\chi \sigma^n\bar\varphi p_n=0.
\ee
With this remark, consistency conditions for the secondary
constraints amount to
\bea\label{lagr}
&& p \l_\L=0, \qquad \L \l_\L=0, \qquad \l_\o=0, \nonumber \\ 
&&2\l_\phi=-i\l_\varphi \s^n \bar\chi p_n
+i\l_\chi \s^n \bar\varphi p_n-i\varphi \s^n \l_{\bar\chi} p_n
+i\chi \s^n \l_{\bar\varphi} p_n, \nonumber \\
&& \L_n {(\s^n \l_{\bar\chi})}_\a+\l_{\L n} 
{(\s^n \bar\chi)}_\a=0,
\qquad
\L_n {(\l_\chi \s^n)}_{\da}+\l_{\L n} {(\chi \s^n)}_{\da}=0,
\nonumber \\
&& \L_n {(\s^n \l_{\bar\varphi})}_\a+\l_{\L n} 
{(\s^n \bar\varphi)}_\a=0,
\qquad
\L_n {(\l_\varphi \s^n)}_{\da}+ \l_{\L n} 
{(\varphi \s^n)}_{\da}=0.
\eea
Making use of the light--cone arguments like those given in the
Appendix  one can show that each of the fermionic equations 
entering Eq. (\ref{lagr}) determines precisely half of the 
corresponding fermionic Lagrange multipliers.

Thus no tertiary constraints arise at this stage, the complete
constraint system being
\be\label{first}
p_e=0, \qquad p_\p=0, \qquad p_{\bar\p}=0,
\ee
\be\label{second1}
p_\r=0,\qquad p_\t-p_n i(\s^n\bar\t)-
\r=0,
\ee
\be\label{second2}
p_{\br}=0, \qquad 
p_{\bar\t}+p_n i(\t{\s}^n)-\br=0, 
\ee
\be\label{second3}
p_\o=0, \qquad \o=0,
\ee
\be\label{second4}
p_\phi=0, \qquad -2\phi-i\varphi \sigma^n \bar\chi p_n
+i\chi \sigma^n\bar\varphi p_n=0,
\ee
\be\label{mixed1}
p_{\varphi}=0, \qquad \varphi \s^n \L_n=0,
\ee
\be\label{mixed2}
p_{\bar\varphi}=0,\qquad \s^n \bar\varphi \L_n=0,
\ee
\be\label{add1}
p_{\chi}=0, \qquad \chi \s^n \L_n=0,
\ee
\be\label{add2}
p_{\bar\chi}=0,\qquad \s^n \bar\chi \L_n=0,
\ee
\be\label{s}
p^2=0, \qquad p_\t \s^n p_n=0, \qquad \s^n p_{\bar\t} p_n=0,
\ee
\be\label{add3}
p_\L=0, \qquad \L^2=0, \qquad 1-\L p=0.  
\ee
The constraints (\ref{first}) are first--class. Imposing the gauge
\be
e =1, \qquad \psi=0, \qquad \bar\psi=0,
\ee
which yields
\be
\lambda_e=0, \qquad \lambda_\psi=0, \qquad \lambda_{\bar\psi}=0,
\ee
one can disregard the canonical pairs $(e,p_e)$, $(\psi,p_\psi)$,
$(\bar\psi,p_{\bar\psi})$. In the same manner, the variables
$(\r,p_\r)$,$(\bar\r,p_{\bar\r})$,$(\o,p_\o)$,$(\phi,p_\phi)$
can be omitted after introducing the Dirac bracket associated 
with the second class constraints 
(\ref{second1})--(\ref{second4}). The Dirac brackets for the 
remaining variables prove to coincide with the Poisson ones. 

One has to be more inventive when imposing a gauge in the sector 
(\ref{mixed1}), ((\ref{mixed2})). Passing to the light--cone 
coordinates (see Appendix) one concludes that, due to $\L^2=0$, there is 
only one linearly 
independent component entering the last of the spinor constraints
(\ref{mixed1}) ((\ref{mixed2})), this proves to be second 
class, whereas the corresponding 
momenta include one first and one second class constraints.
Beautifully enough, on account of the last of the equations 
(\ref{add3}) these can be put into covariant 
(redundant) form
\bea
p_\varphi=0 \Leftrightarrow
\left\{\begin{array}{ll} 
p_\varphi \s^n \L_n=0 & \mbox{first class}\\ 
p_\varphi \s^n p_n=0 & \mbox{second class}
\end{array}
\right. 
\eea 
Fixing a gauge is now obvious (again in a covariant and 
redundant form)
\be\label{g1}
\s^n\varphi p_n=0,
\ee
which yields
\be
\varphi=0,
\ee
when combined with Eq. (\ref{mixed1}). The conservation
in time of the gauge (\ref{g1}) yields
\be
\l_\varphi\s^n p_n=0.
\ee
Together with Eq. (\ref{lagr}) this completely specifies 
$\l_\varphi$. Note also that consistency (${(\varphi)}^{*}=
\bar\varphi$) requires us to impose the complex conjugate 
equation
\be
p_n\s^n\bar\varphi=0  \ \ \rightarrow  \ \  \bar\varphi=0.
\ee
One finally concludes that there is no dynamics in the 
sector $(\varphi,p_\varphi)$, $(\bar\varphi,p_{\bar\varphi})$. 

The same arguments apply to the variables 
$(\chi,p_\chi)$, $(\bar\chi,p_{\bar\chi})$. For our purposes,
however, it is convenient not to impose a gauge in this sector
but rather use these purely auxiliary variables to supplement
Siegel's constraints (\ref{s}) up to irreducible ones.
Actually, it is straightforward to check that the system
(see also Ref.~\cite{sup})
\be\label{fin1}
{\bar\Phi}_{\da} \equiv {(p_\t \s^n p_n+p_{\chi}\s^n \L_n)}_{\da}=0, \qquad
\Phi_\a \equiv {(p_n \s^n p_{\bar\t}+\L_n \s^n p_{\bar\chi})}_\a=0 \qquad \mbox{first class},
\ee
\be\label{fin2}
{\bar\Psi}_{\da} \equiv {(\chi \s^n \L_n+p_{\chi}\s^n p_n)}_{\da}=0, \qquad
\Psi_\a \equiv {(\L_n \s^n \bar\chi+p_n \s^n p_{\bar\chi})}_\a=0 \qquad \mbox{second class}, 
\ee
\be\label{fin2a}
p^2=0 \qquad \mbox{first class},
\ee
is completely equivalent to the initial equations 
(\ref{add1})--(\ref{s}). Here the identities
\bea\label{id1}
{p_\chi}^\a=-{\textstyle{\frac {1}{2\L p}}}
{\bar\Phi}_{\da} 
{({\tilde\s}^m p_m)}^{\da\a}-{\textstyle{\frac {1}{2\L p}}}
{\bar\Psi}_{\da} 
{({\tilde\s}^m \L_m)}^{\da\a}
 -{\textstyle{\frac {1}{2\L p}}}p^2 {p_\t}^\a-
{\textstyle{\frac {1}{2\L p}}}\L^2 \chi^\a,
\eea
\bea\label{id2}
{p_{\bar\chi}}^{\da}=-{\textstyle{\frac {1}{2\L p}}}
{({\tilde\s}^m p_m)}^{\da\a} \Phi_\a- 
{\textstyle{\frac {1}{2\L p}}}{({\tilde\s}^m \L_m)}^{\da\a}
\Psi_\a-{\textstyle{\frac {1}{2\L p}}}p^2 {p_{\bar\t}}^{\da}-
{\textstyle{\frac {1}{2\L p}}}\L^2 {\bar\chi}^{\da},
\eea
prove to be useful. The equivalence just stated implies also that the 
constraint set above is {\it irreducible}, otherwise we would 
have less than $8+1$ equations and Eqs. (\ref{fin1})--(\ref{fin2a}) 
would not be equivalent to (\ref{add1})--(\ref{s}) ($8+1$  
linearly independent components).   

It remains to discuss the bosonic constraints (\ref{add3}). 
Constructing a (weak) projector to the directions orthogonal to the 
vectors $p^n, \L^n$
\be
{\pi_m}^n={\d_m}^n-p_m \L^n-\L_m p^n,
\ee
one can easily extract first class constraints contained in
$p_\L$, the complete constraint set being
\be\label{fin2b}
{\tilde p}_{\L m} \equiv  {(\pi p_\L)}_m=p_{\L m}-(p_\L \L)p_m-(p_\L p)\L_m=0 \qquad \mbox{first class},
\ee
\be\label{fin3}
 p_\L p=0, \qquad \L^2=0, \qquad p_\L \L=0, \qquad 1-\L p=0 
\qquad \mbox{second class}.
\ee
In view of the identities\footnote{Here and in what follows
the symbol $\approx$ means an equality up to
a linear combination of {\it second class} constraints.}
\be\label{ident}
{\tilde p}_\L \L \approx 0, \qquad {\tilde p}_\L p \approx 0,
\ee
one concludes that there are only two linearly independent components 
entering Eq.~(\ref{fin2b}), the total number of 
constraints being sufficient to suppress dynamics in the sector. 
In order to explicitly decouple the first class constraints 
above from the fermionic second class ones 
(\ref{fin2}), it suffices to 
redefine them like
\be
{\tilde p_\L}^n=0 \rightarrow {\tilde p_\L}^n-
{\textstyle{\frac 12}}\chi\s^n{\tilde\s}^m p_\chi p_m-
{\textstyle{\frac 12}}p_{\bar\chi}{\tilde\s}^m \s^n \bar\chi p_m=0.
\ee
As the Dirac bracket associated with the second class 
constraints is introduced, this seems to be inessential here.
 
It is worth mentioning, that the dynamical equivalence of the 
model (24) and the Siegel superparticle~\cite{s1} can be easily 
established if one imposes the non covariant gauge
\be  
\L^i=0, \qquad \mbox{i=1,2}.
\ee

To summarize, in the extended phase space the infinite 
reducibility of the constraints~(\ref{s}) characterizing the
Siegel model can be compensated by that coming 
from the sector of additional 
variables to put the fermionic constraints into an irreducible form.
Residual reducibility proves to fall in the bosonic constraints 
(\ref{fin2b}),(\ref{fin3}). Being the first stage of reducibility, these
admit consistent path integral quantization. 

Quantization of the constraint system 
(\ref{fin1})--(\ref{fin2a}),(\ref{fin2b}),(\ref{fin3}) will be our main 
concern in the next sections.

\vspace{0.4cm}

{\bf 3.3. The Dirac bracket}\\

\vspace{0.4cm}

In the presence of second class constraints both the nilpotency
equation to determine the $BRST$ charge and that to fix
the unitarizing Hamiltonian should be solved under the Dirac 
bracket associated with the full set of second class 
constraints~\cite{bat1}. To construct the latter, it suffices
to convert the matrix of Poisson brackets of second class 
constraints\footnote{The construction proves to be more involved 
when second class constraints in a question are (infinitely) reducible. 
A recipe has been given in Ref.~\cite{henn}}.
Denoting the constraints collectively by 
$\Theta_i=(p_\L p,\L^2,p_\L \L,1-\L p,
{\bar\Psi}_{\da},
{\Psi}_\a)$ and 
$\Gamma_{ij}\equiv\{ \Theta_i,\Theta_j \}$, one finds this to 
be 
\be
\Gamma_{ij}=
\left(
\begin{array}{cccccc}
0 & -2\L p & -p p_\L & p^2 & -{(\chi \s^n p_n)}_{\db} 
& -{(p_n \s^n \bar\chi)}_\b \\
\nonumber\\[2pt]
2\L p & 0 & 2\L^2 & 0 & 0 & 0 \\
\nonumber\\[2pt]
p p_\L & -2\L^2 & 0 & \L p & -{(\chi \s^n \L_n)}_{\db} 
& -{(\L_n \s^n \bar\chi)}_\b \\
\nonumber\\[2pt]
-p^2 & 0 & -\L p & 0 & 0 & 0 \\
\nonumber\\[2pt]
{(\chi \s^n p_n)}_{\da} & 0 & {(\chi \s^n \L_n)}_{\da} & 0 & 
-4{({\tilde\s}^{nm})}_{\da\db} \L_n p_m & 0 \\
\nonumber\\[2pt]
{(p_n \s^n \bar\chi)}_\a & 0 & {(\L_n \s^n \bar\chi)}_\a & 0 
& 0 & -4{(\s^{nm})}_{\a\b} \L_n p_m  \\
\end{array}
\right).
\ee
The corresponding superdeterminant amounts to a simple number coefficient
\be
sdet \ \ \Gamma_{ij}=\textstyle{\frac 14},
\ee
this to be used when constructing the path integral measure in Sec. 5.

Given a supermatrix $F=F_B+F_S$, where $F_B$ and $F_S$ are the
body and the soul respectively~\cite{dewitt}, the inverse supermatrix is
constructed according to the rule~\cite{dewitt} 
\be
F^{-1}={F_B}^{-1}+\sum_{k=1}^{\infty} {(-1)}^k 
{({F_B}^{-1} F_S)}^k {F_B}^{-1}.
\ee
In our case only the first two terms entering the power series 
above prove to be non vanishing, the corresponding inverse 
supermatrix being
\be
\Gamma^{ij}=
\frac {1}{\Delta}
\left(
\begin{array}{cccccc}
0 & \L p & 0 & 2\L^2 & 0 & 0 \\
\nonumber\\[2pt]
-\L p & 0 & p^2 & {p p_\L}^{'} & 
-{\textstyle{\frac 12}}{(\chi \s^n p_n)}^{\db} & 
{\textstyle{\frac 12}}{(p_n \s^n \bar\chi)}^\b \\
\nonumber\\[2pt]
0 & -p^2 & 0 & -2\L p & 0 & 0 \\
\nonumber\\[2pt]
-2\L^2 & -{p p_\L}^{'} & 2\L p & 0 & -{(\chi \s^n \L_n)}^{\db}
& {(\L_n \s^n \bar\chi)}^\b \\
\nonumber\\[2pt]
0 & -{\textstyle{\frac 12}}{(\chi \s^n p_n)}^{\da} & 0 & 
-{(\chi \s^n \L_n)}^{\da} &  
2{({\tilde\s}^{nm})}^{\da\db} \L_n p_m & 0 \\
\nonumber\\[2pt]
0 & {\textstyle{\frac 12}}{(p_n \s^n \bar\chi)}^\a & 0 & 
{(\L_n \s^n \bar\chi)}^\a & 0 &
2{(\s^{nm})}^{\a\b} \L_n p_m  \\
\end{array}
\right),\\
\nonumber \\
\ee
where $\Delta\equiv2({(\L p)}^2-\L^2 p^2)$ and
${p p_\L}^{'}\equiv p p_\L+{\textstyle{\frac 12}}(\chi^2
+{\bar\chi}^2)$.

With the $\Gamma^{ij}$ at hand, the Dirac bracket is 
straightforward to build
\begin{eqnarray}
\{A,B\}_D=\{A,B\}
-\{A,\Theta_i\}\Gamma^{ij}\{\Theta_j,B\}.
\end{eqnarray}
Being rather involved in the general form, the bracket
considerably simplifies  when evaluated in specific coordinate 
sectors\footnote{In what follows we omit the label $D$ attached to the 
Dirac brackets.} (only the brackets to be used below are explicitly 
given here)
\begin{eqnarray}\label{brin}
&&\{\chi^\a,p_{\chi\b}\}={\textstyle{\frac 12}}{\delta^\a}_\b-
{\textstyle{\frac {2}{\Delta}}} \L p
{{(\s_{nm})}_\b}^\a \L^n p^m,\quad
\{\chi^\a,\chi^\b\}={\textstyle{\frac {2}{\Delta}}}p^2
{(\s_{nm})}^{\a\b} \L^n p^m,\nonumber\\
&& \{p_{\chi\a},p_{\chi\b}\}={\textstyle{\frac {2}{\Delta}}}\L^2
{(\s_{nm})}_{\a\b} \L^n p^m; \nonumber\\[2pt]
&&\{\L^n,p_{\L m}\}={\delta^n}_m
-{\textstyle{\frac {2}{\Delta}}}\L p(p^n\L_m+\L^n p_m)+
{\textstyle{\frac {2}{\Delta}}} p^2\L^n\L_m+
{\textstyle{\frac {2}{\Delta}}}\L^2 p^n p_m,\nonumber\\
&&\{\L^n,\L^m \}=0,\quad \{p_{\L n},p_{\L m}\}=
{\textstyle{\frac {2}{\Delta}}}p^2
(\L_n p_{\L m}-\L_m p_{\L n})+{\textstyle{\frac {2}{\Delta}}}p p_\L(p_n\L_m
-\nonumber\\
&&\qquad \qquad \qquad p_m\L_n)+{\textstyle{\frac {2}{\Delta}}}p\L(p_{\L n}p_m-
p_{\L m}p_n)-{\textstyle{\frac {i}{\Delta}}}
(\chi^2-{\bar\chi}^2)\epsilon_{nmkl}\L^k p^l;\nonumber\\[2pt]
&&\{\theta^\a,p_{\theta\b}\}={\delta^\a}_\b,\qquad
\{\theta^\a,\theta^\b\}=0, \qquad  
\{p_{\theta\a},p_{\theta\b}\}=0, \qquad \{p_n,p_m\}=0.
\end{eqnarray}

Analogously, for the cross sectors one finds (in what follows we will not 
need the explicit form of the brackets involving the $x^n$--variable, these
are omitted here)
\begin{eqnarray}\label{brfin}
&& \{p_{\L n},\chi^\a\}=
{\textstyle{\frac {1}{\Delta}}}p^2
\L_n\chi^\a+{\textstyle{\frac {1}{\Delta}}}p^2{(\chi \s_n
{\tilde\s}^k \L_k)}^\a+
{\textstyle{\frac {1}{\Delta}}}p_n
{(\chi \s^k \L_k {\tilde\s}^m p_m)}^\a-\nonumber\\
&& \qquad \qquad \qquad {\textstyle{\frac {1}{\Delta}}}\L p {(\chi \s_n
{\tilde\s}^k p_k)}^\a,\nonumber\\[2pt]
&&\{p_{\L n},p_{\chi\alpha}\}=
{\textstyle{\frac {1}{\Delta}}}\L^2
p_n\chi_\a+{\textstyle{\frac {1}{\Delta}}}\L^2{(\chi \s_n
{\tilde\s}^k p_k)}_\a+
{\textstyle{\frac {1}{\Delta}}}\L_n
{(\chi \s^k p_k {\tilde\s}^m \L_m)}_\a-\nonumber\\
&&\qquad\qquad\qquad {\textstyle{\frac {1}{\Delta}}}\L p 
{(\chi \s_n {\tilde\s}^k \L_k)}_\a.
\end{eqnarray}

Similar relations hold for complex conjugates. In particular,
to derive brackets involving $(\bar\chi, p_{\bar\chi})$ it suffices
to raise $\a$, low $\b$ and then exchange them with dotted indices.

In obtaining Eqs. (\ref{brin}),(\ref{brfin}) the following 
identities
\bea
&& Tr(\s_{ab}\s_{cd})=-{\textstyle{\frac 12}}(\eta_{ac}\eta_{bd}-
\eta_{ad}\eta_{bc})+{\textstyle{\frac i2}}\epsilon_{abcd}\cr
\nonumber\\[2pt]
&& Tr({\tilde\s}_{ab}{\tilde\s}_{cd})=-{\textstyle{\frac 12}}(\eta_{ac}
\eta_{bd}-\eta_{ad}\eta_{bc})-{\textstyle{\frac i2}}\epsilon_{abcd},
\eea
with $\epsilon_{0123}=1$ and  
$\eta_{nm}={\textstyle{diag(-,+,+,+)}}$, prove to be useful.

Finally, it is worth noting that, as long as the path integral 
quantization is concerned, the presence of the $\delta$--function 
of second class constraints in the path integral 
measure~\cite{bat1} allows one to solve the equations on 
the $BRST$--charge and the unitarizing Hamiltonian modulo 
second class constraints~\cite{bat1}. In particular, this will provide further 
simplifications in Eqs. (\ref{brin})--(\ref{brfin}).

\vspace{0.4cm}

{\bf 3.4. The algebra of first class constraints}\\

\vspace{0.4cm}
Having evaluated the Dirac bracket, we are now in a position to  
specify the algebra of the first class constraints (\ref{fin1}), (\ref{fin2a}),
(\ref{fin2b}), the corresponding structure functions to be used 
when constructing the $BRST$ charge.

Taking into account Eq. (\ref{id1}), the identity
\bea
\chi^2=-{\textstyle{\frac {1}{\L p}}}p^2(p_\chi \chi)- 
{\textstyle{\frac {1}{\L p}}} {\bar\Psi}_{\da} 
{({\tilde\s}^m p_m \chi)}^{\da},
\eea
and the fact that according to the general recipe~\cite{bat1} 
it suffices to know the algebra modulo second class constraints, one 
finds the only nontrivial brackets to be
\bea\label{alg}
&&\{ {\tilde p}_{\L n},{\tilde p}_{\L m}\}\approx
{U_{nm}}^k{\tilde p}_{\L k}+
U_{nm}p^2,\qquad
\{ {\tilde p}_{\L n},\Phi_\a \}\approx{U_{n\a}}^\b \Phi_\b+ 
U_{n\a}p^2,\cr
\nonumber\\[2pt]
&&\{ {\tilde p}_{\L n},{\bar\Phi}_{\da} \}
\approx{U_{n\da}}^{\db} {\bar\Phi}_{\db}
+U_{n\da}p^2.
\eea
The explicit form of the structure functions involved is
(${U_{n\da}}^{\db}$, $U_{n\da}$ are obtained by complex conjugation)
\bea
&& {U_{nm}}^k={\textstyle{\frac {2}{\Delta}}}((\L_n p^2-p_n){\delta_m}^k-(\L_m p^2-p_m){\delta_n}^k)\cr
\nonumber\\[2pt]
&& U_{nm}={\textstyle{\frac {i}{\Delta}}}(p_\chi \chi-p_{\bar\chi}\bar\chi)
\epsilon_{nmkl} \L^k p^l,\cr
\nonumber\\[2pt]
&&{U_{n\a}}^\b={\textstyle{\frac 12}}
{{(\s_n {\tilde\s}^k p_k)}_\a}^\b+
{\textstyle{\frac {1}{\Delta}}}\L_n p^2 {\delta_\a}^\b+
{\textstyle{\frac {1}{\Delta}}}(\L_n p^2-p_n) {{(\L^k \s_k {\tilde\s}^l p_l)}_\a}^\b,\cr
\nonumber\\[2pt]
&& U_{n\a}=
{\textstyle{\frac 12}}{(\s_n p_{\bar\t})}_\a-
{\textstyle{\frac {1}{\Delta}}}\L_n{(p^k \s_k p_{\bar\t})}_\a
+{\textstyle{\frac {1}{\Delta}}}(\L_n p^2-p_n) {(\L^k \s_k p_{\bar\t})}_\a.
\eea
Worth noting also are the important algebraic properties of the 
functions obtained (the same holds for complex conjugates)
\bea\label{pr1}
&& U_{nm} \L^m=0, \qquad {U_{n\a}}^\b \L^n \approx 0, \qquad 
U_{n\a} \L^n \approx 0,\nonumber\\
&& U_{nm} p^m=0, \qquad {U_{n\a}}^\b p^n \approx 0, \qquad  U_{n\a} p^n \approx 0.
\eea
These will be of frequent use when establishing 
the nilpotency of the BRST charge in the next section.

\vspace{0.4cm} 

{\bf 4. The BRST charge and the unitarizing Hamiltonian in the minimal ghost sector}

\vspace{0.4cm} 

Proceeding to the $BRST$ quantization, one associates a couple of canonically conjugate ghost variables 
to each of the first class constraints~(\ref{fin1}), (\ref{fin2a}), (\ref{fin2b}) 
$(C^{\dot\alpha},{\bar{\cal P}}_{\dot\alpha})$, $(C^\alpha,{\bar{\cal P}}_\alpha)$,
$(C,\bar{\cal P})$,$(C^n,{\bar{\cal P}}_n)$. The statistics and the ghost number are specified by  
the conventional prescriptions 
\bea 
&& \epsilon (C^A)=\epsilon ({\bar{\cal P}}^A)=\epsilon_A +1, \nonumber\\
&& gh (C^A)=-gh ({\bar{\cal P}}^A)=1.
\eea
To compensate the overcounting in the sector $(C^n,{\bar{\cal P}}_n)$
(only two components entering the bosonic constraint~(\ref{fin2b}) are linearly independent)
one further introduces~\cite{bat1} the secondary ghosts $(C^1,{\bar{\cal P}}^1)$,
$(C^2,{\bar{\cal P}}^2)$ which obey
\begin{eqnarray} 
&& \epsilon (C^{1,2})=\epsilon ({\bar{\cal P}}^{1,2})=0, \nonumber\\
&& gh (C^{1,2})=-gh ({\bar{\cal P}}^{1,2})=2.
\end{eqnarray}
Together with the previously introduced variables these  exhaust the minimal ghost sector for 
the model under consideration.

The $BRST$ charge is defined to be a solution of the nilpotency equation 
\begin{equation}\label{nilp}
\{ \Omega_{min},\Omega_{min} \} \approx0,
\end{equation}
satisfying the boundary condition
\be\label{bc}
\Omega_{min}=\Phi_\a C^\a +{\bar\Phi}_{\da} C^{\da}+{\tilde p}_{\L n} C^n+p^2 C
+{\bar{\cal P}}_n \L^n C^1+{\bar{\cal P}}_n p^n C^2+\dots .
\ee
The first four terms entering Eq.~(\ref{bc}) are typical for the $BRST$ quantization of irreducible
gauge theories. Through Eq.~(\ref{nilp}) they automatically generate the gauge algebra~(\ref{alg}).
The two remaining terms are designed to generate the identities~(\ref{ident}) and are specific to
the treatment of reducible theories.

Calculating the contribution of the boundary terms into Eq.~(\ref{nilp})
\bea\label{nilp1}
&& \{ \Omega_{min},\Omega_{min} \} \approx 
2{\bar{\cal P}}_m \{\Lambda^m,{\tilde p}_{\Lambda n} \} 
C^1 C^n-2({U_{n\alpha}}^\beta \Phi_\beta+U_{n\alpha} 
p^2)C^\alpha C^n- \nonumber \\
&& \quad \quad 2({U_{n\dot\alpha}}^{\dot\beta} 
{\bar\Phi}_{\dot\beta}+U_{n\dot\alpha} p^2) C^{\dot\alpha} 
C^n-({U_{nm}}^k {\tilde p}_{\Lambda k}+U_{nm} p^2) 
C^m C^n+\dots,
\end{eqnarray}
one can partially clarify the structure of the terms which were missing in 
Eq.~(\ref{bc}). In particular, extending the ansatz~(\ref{bc}) 
by means of three new contributions
\bea
{\textstyle{\frac 12}}{\bar{\cal P}}_k {{\tilde U}_{nm}}^k C^m C^n
+{\bar{\cal P}}_\alpha {U_{n\beta}}^\alpha C^\beta C^n+{\bar{\cal P}}_{\dot\alpha} 
{U_{n\dot\beta}}^{\dot\alpha}C^{\dot\beta} C^n, 
\eea
where
\bea
&& {{\tilde U}_{nm}}^k={U_{nm}}^k-{\textstyle{\frac {2}{\Delta}}} p^k 
(\Lambda_n p_m-\Lambda_m p_n),\nonumber\\  
&& {{\tilde U}_{nm}}^k \Lambda^m\approx {\textstyle{\frac {2}{\Delta}}} 
\{ \Lambda^k,p_{\Lambda n} \},\qquad {{\tilde U}_{nm}}^k p^m\approx 0,  
\eea
one can get rid of the first term (which is a manifestation
of the reducibility of the constraints) and those involving
${\tilde p}_\Lambda,\Phi,\bar\Phi$
\bea\label{nilp2}
&& \{ \Omega_{min},\Omega_{min} \}\approx 
-U_{nm}p^2 C^m C^n-2U_{n\alpha} p^2 C^\alpha C^n-2U_{n\dot\alpha} 
p^2 C^{\dot\alpha} C^n-\nonumber\\
&&\quad \quad 2{\bar{\cal P}}_\alpha {U_{n\gamma}}^\alpha 
{U_{m\beta}}^\gamma C^m C^n C^\beta-
2{\bar{\cal P}}_{\dot\alpha} {U_{n\dot\gamma}}^{\dot\alpha} 
{U_{m\dot\beta}}^{\dot\gamma} C^m C^n C^{\dot\beta}+\dots.   
\end{eqnarray}
In order to verify Eq.~(\ref{nilp2}) one has to use the algebraic properties of the 
structure functions~(\ref{pr1}) and the Jacobi identities resulting from the constraint 
algebra
\bea\label{ji1}
&&{{\tilde U}_{[m{\hat a}}}^k {{\tilde U}_{bc]}}^a\approx 0, \qquad
{U_{[m{\hat a}}}^k {U_{bc]}}^a\approx 0,\nonumber\\[2pt] 
&&\{ {\tilde p}_{\L [n},{{\tilde U}_{ab]}}^k \}\approx0, \qquad
\{ {\tilde p}_{\L [n},{U_{ab]}}^k \}\approx0, \nonumber\\[2pt]
&&\{ {\tilde p}_{\L n},{U_{m\a}}^\b \}-\{ {\tilde p}_{\L m},{U_{n\a}}^\b \}
-{U_{nm}}^k {U_{k\a}}^\b \approx0,\nonumber\\[2pt]
&&\{ {\tilde p}_{\L n},U_{m\a} \}-\{ {\tilde p}_{\L m},U_{n\a} \}
-{U_{nm}}^k U_{k\a} \approx0.
\eea
Similar equations hold for complex conjugates. Here the square bracket stands for a
complete antisymmetrization of indices and a hat over an index
means that it is not affected by the antisymmetrization.

It is instructive then to give an explicit form of the terms quadratic in the structure functions
which enter Eq. (\ref{nilp2}) 
\bea
&& {U_{m\a}}^\b {U_{n\b}}^\gamma-{U_{n\a}}^\b {U_{m\b}}^\gamma=\{ {\textstyle{\frac {1}{\Delta}}}
(\L_n p_m-\L_m p_n){{(\L_l \s^l {\tilde\s}^k p_k)}_\a}^\gamma+
{\textstyle{\frac {1}{\Delta}}}(\L_n p_m-\L_m p_n){\delta_\a}^{\gamma}+\nonumber \\
&&\qquad \qquad \qquad {\textstyle{\frac {1}{\Delta}}} \L_m{{(\s_n {\tilde\s}^k p_k)}_\a}^\gamma-
{\textstyle{\frac {1}{\Delta}}} \L_n{{(\s_m {\tilde\s}^k p_k)}_\a}^\gamma-
{\textstyle{\frac {1}{\Delta}}} (\L_m p^2 -p_m){{(\s_n {\tilde\s}^k \L_k)}_\a}^\gamma+\nonumber \\
&&\qquad \qquad \qquad  {\textstyle{\frac {1}{\Delta}}} (\L_n p^2 -p_n){{(\s_m {\tilde\s}^k \L_k)}_\a}^\gamma+
{{(\s_{nm})}_\a}^\b \}p^2 \equiv {\Pi_{mn\a}}^\gamma p^2,\nonumber\\[2pt]
&& \qquad \qquad {\Pi_{mn\a}}^\gamma \L^n\approx0, \qquad {\Pi_{mn\a}}^\gamma p^n\approx0.
\eea
Being factor of $p^2$ this suggests a further amendment to the $\Omega_{min}$
\bea
&& \bar{\cal P} U_{n\a} C^\a C^n+\bar{\cal P} U_{n\da} C^{\da} C^n+
{\textstyle{\frac 12}} \bar{\cal P} U_{nm} C^m C^n- \nonumber \\
&& {\textstyle{\frac 12}} \bar{\cal P} {\bar{\cal P}}_\a  {\Pi_{nm\b}}^\a C^m C^n C^\b-
{\textstyle{\frac 12}} \bar{\cal P} {\bar{\cal P}}_{\da}  {\Pi_{nm\db}}^{\da} C^m C^n C^{\db},
\eea
where ${\Pi_{nm\db}}^{\da}$ is the complex conjugate of ${\Pi_{nm\b}}^{\a}$.

Beautifully enough, by making use of the next portion of the Jacobi identities
\bea\label{ji7}
&&\{ {\tilde p}_{\L [a},U_{bc]} \}+U_{[a{\hat d}} {U_{bc]}}^d \approx0,\nonumber\\[2pt]
&&{\Pi_{mn\a}}^\gamma \Phi_\gamma+{U_{m\a}}^\b U_{n\b}-{U_{n\a}}^\b U_{m\b}+
\{ \Phi_\a,U_{nm} \}\approx0, \nonumber\\[2pt] 
&&\{ {\tilde p}_{\L [a},{\Pi_{mn]\b}}^\a \}+{\Pi_{[a{\hat d}\b}}^\a 
{U_{mn]}}^d \approx0,\nonumber\\[2pt]
&&{\Pi_{[mn\delta}}^\gamma {U_{a]\gamma}}^\a-
{U_{[a\delta}}^\gamma {\Pi_{mn]\gamma}}^\a\approx0, 
\eea
and their complex conjugates, one can verify the nilpotency of our ansatz, 
the $BRST$ charge in the minimal ghost sector being of the form
\bea\label{omegamin}
&& \Omega_{min}=\Phi_\a C^\a +{\bar\Phi}_{\da} C^{\da}+{\tilde p}_{\L n} C^n+p^2 C
+{\bar{\cal P}}_n \L^n C^1+{\bar{\cal P}}_n p^n C^2+ \nonumber\\
&& \qquad \qquad {\textstyle{\frac 12}}{\bar{\cal P}}_k {{\tilde U}_{nm}}^k C^m C^n
+{\bar{\cal P}}_\a {U_{n\b}}^\a C^\b C^n+{\bar{\cal P}}_{\da} {U_{n\db}}^{\da}
C^{\db} C^n+\nonumber\\   
&& \qquad \qquad \bar{\cal P} U_{n\a} C^\a C^n+\bar{\cal P} U_{n\da} C^{\da} C^n+
{\textstyle{\frac 12}} \bar{\cal P} U_{nm} C^m C^n- \nonumber \\
&& \qquad \qquad {\textstyle{\frac 12}} \bar{\cal P} {\bar{\cal P}}_\a  {\Pi_{nm\b}}^\a C^m C^n C^\b-
{\textstyle{\frac 12}} \bar{\cal P} {\bar{\cal P}}_{\da}  {\Pi_{nm\db}}^{\da} C^m C^n C^{\db}.
\eea
For this to be real, one has to impose the following conjugation properties
on the ghost variables
\bea
&& {(C^\a)}^{*}=C^{\da}, \quad {(C^n)}^{*}=C^n, \quad {(C)}^{*}=C,\quad
{(C^{1,2})}^{*}=-C^{1,2}, \nonumber \\
&& {({\bar{\cal P}}_\a)}^{*}={\bar{\cal P}}_{\da}, \quad 
{({\bar{\cal P}}_n)}^{*}=-{\bar{\cal P}}_n, 
\quad {(\bar{\cal P})}^{*}=-\bar{\cal P}, \quad
{({\bar{\cal P}}^{1,2})}^{*}=-{\bar{\cal P}}^{1,2}.
\eea

Thus, within the framework of the $BRST$ quantization the modified formulation proves to be a 
theory of rank two. Our result here correlates well with that obtained previously in the alternative 
harmonic superspace approach~\cite{nis}. Worth noting also is that a naive limit of the expression obtained 
to the original phase space  breaks manifest
Lorenz covariance $(\L^i=0, \L^{-}=0, \L^{+}=\textstyle{-\frac{1}{p^{-}}})$, as it should.

Finally, we observe that the boundary condition which has to be imposed on the unitarizing 
Hamiltonian (a proper Hamiltonian treatment requires secondary constraints to be added to
the initial Hamiltonian with the corresponding Lagrange multipliers)
\be\label{unham}
H |_{C=\bar{\cal P}=0}=H_0=0,
\ee
automatically satisfies the needed equation
\be
\{H,\Omega_{min}\}\approx 0.
\ee
Hence, no ghost corrections to Eq.~(\ref{unham}) are to be added, the latter fits to describe 
the unitarizing Hamiltonian for the case at hand (see also Ref.~\cite{diaz}).

\vspace{0.4cm} 

{\bf 5. Extension to the nonminimal ghost sector. Transition amplitude.}

\vspace{0.4cm} 

Having constructed $\Omega_{min}$ and $H$, an extension to the nonminimal ghost sector is 
straightforward~\cite{bat1}. The irreducible constraints $\Phi_\a$, ${\bar\Phi}_{\da}$, $p^2$ can be 
treated in the usual way. One introduces three canonical pairs of new 
ghost variables 
$({\cal P}^\a,{\bar C}_\a)$,$({\cal P}^{\da},{\bar C}_{\da})$, 
$({\cal P}, \bar C)$ along with the corresponding Lagrange multipliers
$(\l^\a,\pi_\a)$,$(\l^{\da},\pi_{\da})$,$(\l,\pi)$ (the statistics and the ghost number of the new 
variables are given below in the Table 1). Associated with the reducible constraints ${\tilde p}_\L$ 
are the primary ghosts and Lagrange multipliers $({\cal P}^n,{\bar C}_n)$, $(\l^n,\pi_n)$, as well 
as the secondary ones~\cite{bat1} $({\cal P}^1,{\bar C}^1)$, $({\cal P}^2,{\bar C}^2)$, $(\l^1,\pi^1)$, 
$(\l^2,\pi^2)$. A direct inspection of 
the structure of the ghost sector (with the use of the Table 1) shows the 
disbalance between the 
number of unphysical degrees of freedom and that of the ghosts introduced. This can be improved by 
introducing further ``extra'' ghosts~\cite{bat1}. In our case these are exhausted by  
$({{\cal P}_{(1)}}^1,{{\bar C}_{(1)}}^1)$, $({{\cal P}_{(1)}}^2,
{{\bar C}_{(1)}}^2)$, 
$({\l_{(1)}}^1,{\pi_{(1)}}^1)$, $({\l_{(1)}}^2,{\pi_{(1)}}^2)$. The statistics and the ghost number of 
the new variables are gathered in the following 
table
\vspace{0.3cm}
\begin{center}
Table 1.1 Ghosts (nonminimal sector)
\end{center}
\begin{center}
\begin{tabular}{|l|c|c|c|c|c|c|c|c|c|c|c|c|}
\hline		\vphantom{$\displaystyle\int$}
  & ${\cal P}^\a$ & ${\bar C}_\a$ & ${\cal P}^{\da} $ & ${\bar C}_{\da}$ 
& $\cal P $ & $\bar C$ & ${\cal P}^n$ & ${\bar C}_n$ & ${\cal P}^1$ &
${\bar C}^1$ & ${\cal P}^2$ & ${\bar C}^2$ \\
\hline		\vphantom{$\displaystyle\int$}
$\epsilon$ & 0 & 0 & 0 & 0 
	& 1 & 1 & 1 & 1 & 0 & 0 & 0 & 0\\
\hline		\vphantom{$\displaystyle\int$}
$gh$ & 1 & -1 & 1 & -1 & 1 & -1 & 1 & -1 & 2 & -2 & 2 & -2 \\
\hline
\end{tabular}
\end{center}

\vspace{0.2cm}

\begin{center}
Table 1.2 Lagrange multipliers
\end{center}
\begin{center}
\begin{tabular}{|l|c|c|c|c|c|c|c|c|c|c|c|c|}
\hline		\vphantom{$\displaystyle\int$}
  & $\l^\a$ & $\pi_\a$ & $\l^{\da} $ & $\pi_{\da}$ 
& $\l $ & $\pi$ & $\l^n$ & $\pi_n$ & $\l^1$ &
$\pi^1$ & $\l^2$ & $\pi^2$ \\
\hline		\vphantom{$\displaystyle\int$}
$\epsilon$ & 1 & 1 & 1 & 1 
	& 0 & 0 & 0 & 0 & 1 & 1 & 1 & 1\\
\hline		\vphantom{$\displaystyle\int$}
$gh$ & 0 & 0 & 0 & 0 & 0 & 0 & 0 & 0 & 1 & -1 & 1 & -1 \\
\hline
\end{tabular}
\end{center}

\vspace{0.2cm}

\begin{center}
Table 1.3 Extra ghosts
\end{center}
\begin{center}
\begin{tabular}{|l|c|c|c|c|c|c|c|c|}
\hline		\vphantom{$\displaystyle\int$}
 & ${{\cal P}_{(1)}}^1$ & ${{\bar C}_{(1)}}^1$ & ${{\cal P}_{(1)}}^2 $ & 
${{\bar C}_{(1)}}^2$ & ${\l_{(1)}}^1$ & ${\pi_{(1)}}^1$& ${\l_{(1)}}^2$ 
& ${\pi_{(1)}}^2$ \\
\hline		\vphantom{$\displaystyle\int$}
$\epsilon$ & 1 & 1 & 1 & 1 
	& 0 & 0 & 0 & 0\\
\hline		\vphantom{$\displaystyle\int$}
$gh$ & 1 & -1 & 1 & -1 & 0 & 0 & 0 & 0 \\
\hline
\end{tabular}
\end{center}

A continuation of $\Omega_{min}$ to the complete relativistic phase space is now easy to perform
\bea\label{nonmin}
&& \Omega=\Omega_{min}+\pi_\a {\cal P}^\a+\pi_{\da} {\cal P}^{\da}+\pi {\cal P}+\pi_n {\cal P}^n
+\pi^1 {\cal P}^1+\pi^2 {\cal P}^2+\nonumber\\
&& \quad \quad {\pi_{(1)}}^1 {{\cal P}_{(1)}}^1+{\pi_{(1)}}^2 {{\cal P}_{(1)}}^2.
\eea
This supplies us with the last tool needed for quantizing the theory, the corresponding transition
amplitude is given by the formal path integral\footnote{As usual, the fermionic $\delta$--function
is defined as $\delta(\t)=\t$. Hence, $\delta({\Psi_\a})\sim \Psi_1 \Psi_2 \sim \Psi^2$.}
\be\label{trans}
Z_\Psi={\textstyle{\frac 12}} \int D\mu \delta({\L^2})\delta(1-\L p)\delta(p_\L p)
\delta(p_\L \L)\delta(\Psi_\a)\delta({\bar\Psi}_{\da})
e^{{\frac ih}S}.
\ee
Here the effective quantum action has the form
\bea\label{eff}
&& S=\int d\tau (p_n {\dot x}^n+p_{\t\a}{\dot\t}^\a+{p_{\bar\t}}^{\da}{\dot{\bar\t}}_{\da}+
p_{\L n} {\dot\L}^n+p_{\chi\a}{\dot\chi}^\a+{p_{\bar\chi}}^{\da}{\dot{\bar\chi}}_{\da}+\pi_\a {\dot\l}^\a+
\pi^{\da}{\dot\l}_{\da}+\nonumber\\
&& \quad \quad \pi\dot\l+\pi_n {\dot\l}^n+\pi^1{\dot\l}^1+\pi^2{\dot\l}^2
+{\pi_{(1)}}^1{{\dot\l}_{(1)}}^1+{\pi_{(1)}}^2 {{\dot\l}_{(1)}}^2+{\bar{\cal P}}_\a {\dot C}^\a+
{\bar{\cal P}}_{\da} {\dot C}^{\da}+\nonumber\\
&& \quad \quad {\bar{\cal P}} {\dot C}+{\bar{\cal P}}_n {\dot C}^n+
{\bar{\cal P}}^1 {\dot C}^1+{\bar{\cal P}}^2 {\dot C}^2+{\bar C}_\a {\dot{\cal P}}^\a
+{\bar C}_{\da} {\dot{\cal P}}^{\da}+{\bar C} \dot{\cal P}+{\bar C}_n {\dot{\cal P}}^n+\nonumber\\
&& \quad \quad {\bar C}^1 {\dot{\cal P}}^1+{\bar C}^2 {\dot{\cal P}}^2+
{\bar C_{(1)}}^1 {\dot{\cal P}_{(1)}}^1+{\bar C_{(1)}}^2 {\dot{\cal P}_{(1)}}^2-\{ \Psi,\Omega \}_D),
\eea
with $D\mu$ being the usual Liouville measure over the full phase space and $\Psi$ denoting the 
gauge fixing fermion $(\epsilon(\Psi)=1, gh(\Psi)=-1)$. Given a specific form of the latter,
a number of ghost (and Lagrange multiplier) integrations can be
performed explicitly. For example, we can integrate out the whole bunch of variables $(C^{1,2},{\bar {\cal P}}^{1,2})$,
$({\cal P}^{1,2},{\bar C}^{1,2})$,  $(\l^{1,2},\pi^{1,2})$, $({{\cal P}_{(1)}}^{1,2},{{\bar C}_{(1)}}^{1,2})$, 
$({\l_{(1)}}^{1,2},{\pi_{(1)}}^{1,2})$ by taking the following representation for $\Psi$ 
\be
\Psi=\l^1 {\bar {\cal P}}^1+\l^2 {\bar {\cal P}}^2+{\textstyle{\frac {1}{\epsilon}}}C_n p^n {\bar C}^1+
{\textstyle{\frac {1}{\epsilon}}}C_n \L^n {\bar C}^2+{\textstyle{\frac {1}{\epsilon}}}{\l_{(1)}}^1 
{{\bar C}_{(1)}}^1+{\textstyle{\frac {1}{\epsilon}}}{\l_{(1)}}^2 
{{\bar C}_{(1)}}^2+\Psi^{'}.
\ee 
Here $\Psi^{'}$ does not depend on the set above and $\epsilon$ is a constant. An explicit integration which appeals to a
passage to a discrete lattice attaches then four new factors to the path integral measure 
\bea\label{newmeasure}
&& Z_\Psi={\textstyle{\frac 12}} \int D\mu^{'} \delta({\L^2})\delta(1-\L p)\delta(p_\L p)
\delta(p_\L \L)\delta(\Psi_\a)\delta({\bar\Psi}_{\da})\nonumber\\
&& \qquad \qquad\delta (C_n p^n)
\delta (C_n \L^n)\delta ({\cal P}_n p^n)\delta ({\cal P}_n \L^n)e^{{\frac ih}S},
\eea
and reduces the effective action to the relatively simple form
\bea\label{eff}
&& S=\int d\tau (p_n {\dot x}^n+p_{\t\a}{\dot\t}^\a+{p_{\bar\t}}^{\da}{\dot{\bar\t}}_{\da}+
p_{\L n} {\dot\L}^n+p_{\chi\a}{\dot\chi}^\a+{p_{\bar\chi}}^{\da}{\dot{\bar\chi}}_{\da}+\pi_\a {\dot\l}^\a+
\pi^{\da}{\dot\l}_{\da}+\nonumber\\
&& \quad \quad \pi\dot\l+\pi_n {\dot\l}^n+{\bar{\cal P}}_\a {\dot C}^\a+
{\bar{\cal P}}_{\da} {\dot C}^{\da}+{\bar{\cal P}} {\dot C}+{\bar{\cal P}}_n {\dot C}^n+{\bar C}_\a {\dot{\cal P}}^\a
+{\bar C}_{\da} {\dot{\cal P}}^{\da}+{\bar C} \dot{\cal P}\nonumber\\
&&\quad \quad +{\bar C}_n {\dot{\cal P}}^n-\{ \Psi^{'},\Omega^{'}\}_D),
\eea
where $\Omega^{'}$ is given by Eq. (\ref{nonmin}) with the terms involving $C^{1,2}, \pi^{1,2}, {\pi_{(1)}}^{1,2}$
omitted. In the course of the integration the standard change of variables (with unit Jacobian)
\be
\pi^{1,2} \rightarrow \epsilon \pi^{1,2}, \qquad {\bar C}^{1,2} \rightarrow \epsilon {\bar C}^{1,2},
\ee 
followed by the limit $\epsilon \rightarrow 0$ has been used. Note that we do not see the compensating 
ghosts $(C^{1,2}, {\bar {\cal P}}^{1,2})$ any more. The overcounting intrinsic to the sector 
$(C^n, {\bar {\cal P}}_n)$ is regulated now by the measure in Eq. (\ref{newmeasure}). 
 
In general, we can proceed on this way. However, this seems to break manifest Lorentz covariance.
In particular we found that the following ansatz for $\Psi^{'}$
\bea
&& \Psi^{'}={\textstyle{\frac {1}{\epsilon}}} \l^{+} {\bar C}^{-}+
{\textstyle{\frac {1}{\epsilon}}} \l^{-} {\bar C}^{+}+{\textstyle{\frac {1}{\epsilon}}}  \L_i {\bar C}^i  
+{\textstyle{\frac {1}{\epsilon}}} {\bar C}_{0} {\bar \chi}^{\dot 1} 
+{\textstyle{\frac {1}{\epsilon}}} {\bar C}_{\dot 0} \chi^{1} 
+{\bar {\cal P}}^i \l_i\nonumber\\
&&+({\bar {\cal P}}_{0}-{\textstyle{\frac{p^{1}-i p^{2}}{\sqrt{2} p^{-}}}} {\bar {\cal P}}_{1}) \l^{0}+
({\bar{\cal P}}_{\dot 0}-{\textstyle{\frac{p^{1}+i p^{2}}{\sqrt{2} p^{-}}}} {\bar{\cal P}}_{\dot 1}) \l^{\dot 0}+
\Psi^{''},
\eea
where $\Psi^{''}$ depends on $(p^n, \t^\a, {p_\t}_\a, {\bar\t}^{\dot\a}, {p_{\bar\t}}_{\dot\a})$ only 
and we switch to the light--cone notation $\l^n \rightarrow (\l^{\pm},\l^i)$ (for fermions we 
write the indices explicitly
${\bar C}_\a=({\bar C}_{0}, {\bar C}_{1})$), reduces the 
integral to the standard path integral constructed with respect to the irreducible (noncovariant) subset of
the Siegel constraints (\ref{sconst})
\be
p_{\t 0}-{\textstyle{\frac{p^1-ip^2}{\sqrt{2} p^{-}}}} p_{\t 1}=0, \qquad
p_{ \bar\t \dot 0}-{\textstyle{\frac{p^1+ip^2}{\sqrt{2} p^{-}}}} 
p_{ \bar\t \dot 1}=0, \qquad
p^2=0.
\ee
The explicit form of the latter is (we denote collectively $z=(x^n,\t^1,{\bar\t}^{\dot 1})$)
\be\label{noncovprop}
K(z_f, t_f;z_i,t_i)=\delta({\t^1}_{f}-{\t^1}_{i}) 
\delta({{\bar\t}^{\dot 1}}_{f}-{{\bar\t}^{\dot 1}}_{i}) 
K_{0} (x_f,t_f;x_i,t_i)
\ee
where $K_{0} (x_f,t_f;x_i,t_i)$ is the propagator of the massless spinless relativistic particle
and $({\t^1}_{i},{{\bar\t}^{\dot 1}}_{i},{\t^1}_{f},{{\bar\t}^{\dot 1}}_{f})$ denote the values of the 
fermions at the initial and final moments of time (boundary conditions).

Thus the path integral constructed above can be viewed as a formal covariantization of
the propagator (\ref{noncovprop}) characterizing the Siegel superparticle. Beautifully enough, 
this can be done with a finite number of ghost variables.

\vspace{0.4cm} 

{\bf 6. Discussion}

\vspace{0.4cm} 
In this article we have studied an alternative to the harmonic superspace approach,
the latter seems to be the only method for quantizing infinitely reducible first class
constraints currently available. The basic advantage of the novel technique is the
existence of an explicit Lagrangian formulation and the validity of the standard spin--statistics relations
for all the variables involved. In contrast to the harmonic superspace approach, where one 
first extracts linearly independent components from originally reducible constraints and then
quantizes the resulting irreducible theory, the infinite reducibility of constraints is
effectively canceled by that coming from the sector of auxiliary variables. Both methods, however,
correlate well yielding  a theory of rank two after $BRST$ quantization. 

Turning to possible further developments, one expects the treatment of the $ABCD$ model along 
similar lines to be a natural next step. As has been mentioned in the Introduction, however, a proof 
of the equivalence of the $ABCD$ superparticle to a conventional model does not treat all constraints 
on equal footing. In view of this fact, the stringy extension seems to be  
preferable. Then, as was recently  marked by Berkovits~\cite{ber}, a naive generalization of the present 
scheme to the superstring case~\cite{der} faces the zero mode problem and, hence, deserves further 
investigation. We suspect, however, the latter point to be a technical difficulty rather than 
an ideological one. Another interesting point is to make use of the present approach to test an 
earlier quantization proposal by Kallosh~\cite{kal2} (see also related works~\cite{diaz},\cite{bell1}). 
The infinite proliferation of ghosts has been truncated there by
imposing appropriate conditions on the ghosts variables, the latter involving
specific (covariant) projectors. The phase space in our method is valid for the construction of such projectors
(see also Ref.~\cite{diaz}) and the possibility to truncate the infinite ghost tower following Kallosh's approach at the 
very second step seems to be tempting.

\vspace{0.4cm}

{\bf Acknowledgments}

\vspace{0.4cm} 

A.G. thanks Igor Bandos for useful discussions.

\vspace{0.4cm}

{\bf Appendix}

\vspace{0.4cm}

In this Appendix we prove the equivalence of the last of 
Eqs.~(\ref{secondary}) and the pair (\ref{o}), provided other 
constraints from~(\ref{secondary}) hold. Some details related to 
the analysis of the constraint system in the light--cone frame
are also given.

Given the vector equation
$$
\arr
-2\phi\L^n+\o p^n-i\varphi \sigma^n \bar\chi
+i\chi \sigma^n\bar\varphi=0,
\ea
\eqno{(A.1)}
$$
the multiplication by $\L_n$ gives
$$
\arr
\o=0.
\ea
\eqno{(A.2)}
$$
Hence, the second term in $(A.1)$ can be omitted. Passing to 
light cone coordinates, one has 
$$
\arr
-2\phi\L^{+}-i\varphi \sigma^{+} \bar\chi
+i\chi \sigma^{+}\bar\varphi=0,
\ea
\eqno{(A.3)}
$$
$$
\arr
-2\phi\L^{-}-i\varphi \sigma^{-} \bar\chi
+i\chi \sigma^{-}\bar\varphi=0,
\ea
\eqno{(A.4)}
$$
$$
\arr
-2\phi\L^{i}-i\varphi \sigma^{i} \bar\chi
+i\chi \sigma^{i}\bar\varphi=0,
\ea
\eqno{(A.5)}
$$
where the customary notation 
$\L^{\pm}=\pm{\textstyle{\frac {1}{\sqrt{2}}}}(\L^{0}\pm\L^{3})$
is used.

It is worth mentioning now that, given a light--like vector 
$\L^2=-2\L^{+}\L^{-}+\L^{i}\L^{i}=0$, the equation 
${(\varphi \s^n)}_{\da} \L_n=0$ contains only half (one) 
linearly independent components. Actually, taking a
conventional set of $\s$--matrices in ${R}^{1|3}$ (see Ref.~\cite{kuz}) 
$$
\begin{array}{lll}
\{\sigma_n,{\tilde\sigma}_m\}=-2\eta_{nm}, \qquad
\eta_{nm}={\textstyle{diag(-,+,+,+)}},\nonumber\\[2pt]

\s^{+}=-{\textstyle{\sqrt{2}}}\left(\begin{array}{cc} 0 & 0\\
0 & 1\end{array}\right), \qquad
\s^{-}=-{\textstyle{\sqrt{2}}}\left(\begin{array}{cc} 1 & 0\\
0 & 0 \end{array}\right),\nonumber\\[2pt]

\L_n \s^n={\textstyle{\sqrt{2}}}\left(\begin{array}{cc} 
{\textstyle{\L^{+}}} & 
{\textstyle{\frac{\L^{1}-i\L^{2}}{\sqrt{2}}}}\\
{\textstyle{\frac{\L^{1}+i\L^{2}}{\sqrt{2}}}}
& {\textstyle{\L^{-}}} \end{array}\right), \qquad

\L_n {\tilde\s}^n={\textstyle{\sqrt{2}}}\left(\begin{array}{cc} 
{\textstyle{\L^{-}}} & 
-{\textstyle{\frac{\L^{1}-i\L^{2}}{\sqrt{2}}}}\\
-{\textstyle{\frac{\L^{1}+i\L^{2}}{\sqrt{2}}}}
& {\textstyle{\L^{+}}} \end{array}\right), 
\end{array}
\eqno{(A.6)}
$$
with ${\tilde\sigma}^{n\dot\alpha \alpha}=
\epsilon^{\dot\alpha\dot\beta}\epsilon^{\alpha\beta}
\sigma_{n\beta\dot\beta}$, one finds
$$
\arr
{(\varphi \s^n)}_{\da} \L_n=0 \Rightarrow
\left\{\begin{array}{ll} 
\varphi^{0} \L^{+}+
\varphi^{1}{\textstyle{\frac{(\L^{1}+i\L^{2})}{\sqrt{2}}}} =0 &\\ 
\varphi^{0}{\textstyle{\frac{(\L^{1}-i\L^{2})}{\sqrt{2}}}}+
\varphi^{1} \L^{-}=0. &\\ 
\end{array}
\right.
\ea
\eqno{(A.7)}
$$
Multiplying the first equation in $(A.7)$ by 
${\textstyle{\frac{(\L^{1}-i\L^{2})}{\sqrt{2}}}}$ one recovers 
the second one, provided the standard light--cone condition
$$
\arr
\L^{+}\neq 0
\ea
\eqno{(A.8)}
$$
is assumed. 

With the use of the explicit representation of the $\s$--matrices
chosen, the constraint system $(A.3)$--$(A.5)$ simplifies to 
$$
\arr
-2\phi\L^{+}+i\sqrt{2}\varphi^{1}{\bar\chi}^{\dot{1}}
-i\sqrt{2}\chi^{1}{\bar\varphi}^{\dot{1}}=0,
\ea
\eqno{(A.9)}
$$
$$
\arr
-2\phi\L^{-}+i\sqrt{2}\varphi^{0}{\bar\chi}^{\dot{0}}
-i\sqrt{2}\chi^{0}{\bar\varphi}^{\dot{0}}=0,
\ea
\eqno{(A.10)}
$$
$$
\arr
-\phi(\L^{1}+i\L^{2})-i\varphi^{0}{\bar\chi}^{\dot{1}}
+i\chi^{0}{\bar\varphi}^{\dot{1}}=0,
\ea
\eqno{(A.11)}
$$
$$
\arr
-\phi(\L^{1}-i\L^{2})-i\varphi^{1}{\bar\chi}^{\dot{0}}
+i\chi^{1}{\bar\varphi}^{\dot{0}}=0.
\ea
\eqno{(A.12)}
$$
On account of Eq. $(A.7)$ (the same holds for $\chi$ and  
complex conjugates), the last three equations  
follow from $(A.9)$. Thus, there appears to be only one 
linearly independent component entering the original vector 
equation. The latter can be put into a covariant 
(scalar) form. Actually, applying the same light--cone 
technique to the equation 
$$
\arr
-2\phi-i\varphi \sigma^n \bar\chi p_n
+i\chi \sigma^n\bar\varphi p_n=0,
\ea
\eqno{(A.12)}
$$
one recovers precisely Eq. $(A.9)$.

\end{document}